\documentclass[a4paper,12pt]{article}
\usepackage{amssymb}
\usepackage{mathrsfs}
\usepackage{bbm,dcolumn,bm}
\usepackage{epsf,amsmath,amssymb}
\usepackage[dvips,usenames]{color}
\usepackage{graphicx}
\usepackage{color}
\usepackage{colortbl}

\newlength{\dinwidth}
\newlength{\dinmargin}
\def\pslash{\rlap{\hspace{0.02cm}/}{p}}
\def\kslash{\rlap{\hspace{0.02cm}/}{k}}
\newcommand{\half}{\frac{1}{2}}

\begin{document}
\title{\Large \bf The effects of a family non-universal $Z^{\prime}$ boson on $B\to \pi\pi$ decays}
\author{Qin Chang$^{a,b}$\footnote{E-mail: changqin@htu.cn}, Xin-Qiang Li$^{a,c}$\footnote{E-mail: xqli@itp.ac.cn}, Ya-Dong Yang$^{b,d}$\footnote{E-mail: yangyd@iopp.ccnu.edu.cn}\\
{ $^a$\small Department of Physics, Henan Normal University, Xinxiang, Henan 453007, P.~R. China}\\
{ $^b$\small Institute of Particle Physics, Huazhong Normal University, Wuhan, Hubei  430079, P.~R. China}\\
{ $^c$\small IFIC, Universitat de Val\`encia-CSIC, Apt. Correus 22085, E-46071 Val\`encia, Spain}\\
{ $^d$\small Key Laboratory of Quark \& Lepton Physics, Ministry of Education, Huazhong Normal University,}\\ {\small Wuhan, Hubei, 430079, P.~R. China}}

\date{}
\maketitle
\bigskip\bigskip
\maketitle \vspace{-1.5cm}

\begin{abstract}
{\noindent} Motivated by the measured large branching ratio of $\bar{B}^{0}\to\pi^0\pi^0$~(the so-called ``$\pi\pi$'' puzzle), we investigate the effects of a family non-universal $Z^{\prime}$ model on the tree-dominated $B\to\pi\pi$ decays. We find that the $Z^{\prime}$ coupling parameter $\zeta_{d}^{LR}\sim0.05$ with a nontrivial new weak phase $\phi_d^L\sim-50^{\circ}$, which is relevant to the $Z^{\prime}$ contributions to the QCD penguin sector $\triangle C_5$, is needed to reconcile the observed discrepancy. Combined with the recent fitting results from $B\to\pi K$, $\pi K^{\ast}$ and $\rho K$ decays, the $Z^{\prime}$ parameter spaces are severely reduced but still not excluded entirely, implying that both the ``$\pi\pi$'' and ``$\pi K$'' puzzles could be accommodated simultaneously within such a family non-universal $Z^{\prime}$ model.
\end{abstract}


\newpage

\section{Introduction}
\label{sec:intro}

With the fruitful running of both BaBar and Belle in the past decade, as well as the upcoming LHC-b and the proposed Super-B experiments, rare B-meson decays provide a golden opportunity to test the Standard Model~(SM) picture of flavor physics and CP violation. Although most experimental measurements are in perfect agreement with the SM predictions, some incomprehensible discrepancies, such as the measured large branching ratio of $\bar{B}^{0}\to\pi^0\pi^0$~\cite{HFAG}~(the so-called ``$\pi\pi$ puzzle'') and the direct CP asymmetries $A_{CP}(B^-\to \pi^0 K^-)\neq A_{CP}(\bar{B}^0\to\pi^+ K^-)$ at $5\sigma$ significance~\cite{:2008zza}~(the so-called ``$\pi K$ puzzle''~\cite{pipipuz1,pipipuz2}), still exist.

The observed ``$\pi\pi$ puzzle'' is reflected by the following two ratios of CP-averaged branching
fractions~\cite{pipipuz1,pipipuz2}
%%%%%%%%%%%%%%%%%%%%%%%%%%%%%%%%%%%%%%%%%%%%%%%%%
\begin{eqnarray}
 R_{+-}^{\pi\pi}\equiv2\Big[\frac{{\cal B}(B^{-}\to\pi^{-}\pi^0)}{{\cal
 B}(\bar{B}^{0}\to\pi^{+}\pi^{-})}\Big]\frac{\tau_{B^0}}{\tau_{B^+}}\,,\qquad
 R_{00}^{\pi\pi}\equiv2\Big[\frac{{\cal B}(\bar{B}^{0}\to\pi^0\pi^0)}{{\cal
 B}(\bar{B}^{0}\to\pi^{+}\pi^{-})}\Big]\,.
\end{eqnarray}
%%%%%%%%%%%%%%%%%%%%%%%%%%%%%%%%%%%%%%%%%%%%%%%%%
With the up-to-date results averaged by the Heavy Flavor Averaging Group~(HFAG)~\cite{HFAG} from BaBar \cite{BaBar1,BaBar2,BaBar3}, Belle~\cite{Belle1,Belle2}, CLEO~\cite{CLEO} and CDF~\cite{CDF}, and taking $\tau_{B^+}/\tau_{B^0}=1.071\pm0.009$~\cite{PDG08}, we get
%%%%%%%%%%%%%%%%%%%%%%%%%%%%%%%%%%%%%%%%%%%%%%%%%
\begin{eqnarray}\label{expR}
 R_{+-}^{\pi\pi}(\text{Exp.})=2.02\pm0.17\,,\qquad
 R_{00}^{\pi\pi}(\text{Exp.})=0.60\pm0.08\,.
\end{eqnarray}
%%%%%%%%%%%%%%%%%%%%%%%%%%%%%%%%%%%%%%%%%%%%%%%%%
Theoretically, it is generally expected that ${\cal B}(\bar{B}^{0}\to\pi^+\pi^-)>{\cal B}(B^{-} \to\pi^{-}\pi^0)$ and ${\cal B}(\bar{B}^{0}\to\pi^+\pi^-)\gg{\cal B}(\bar{B}^{0}\to\pi^0\pi^0)$ within the  SM. Adopting the central values calculated within the QCD factorization~(QCDF)~\cite{Beneke11,Beneke12,Beneke13}, the  perturbative QCD~(pQCD)~\cite{KLS1,KLS2,KLS3} and the soft-collinear theory~(SCET)~\cite{scet1,scet2,scet3,scet4}, we get respectively
%%%%%%%%%%%%%%%%%%%%%%%%%%%%%%%%%%%%%%%%%%%%%%%%%
\begin{eqnarray}
\begin{aligned}
 R_{+-}^{\pi\pi}(\text{QCDF})&=1.25, 1.83\,(\text{S4})\,,\qquad
 &&R_{00}^{\pi\pi}(\text{QCDF})=0.07, 0.27\,(\text{S4})\,\cite{Beneke3}\,,\\
 R_{+-}^{\pi\pi}(\text{pQCD})&=0.93, 1.15\,(\text{+NLO})\,,\qquad
 &&R_{00}^{\pi\pi}(\text{pQCD})=0.03, 0.09\,(\text{+NLO})\,\cite{PiPipQCD}\,,\\
 R_{+-}^{\pi\pi}(\text{SCET})&=1.80,\qquad
 &&R_{00}^{\pi\pi}(\text{SCET})=0.31\,\cite{PiPiSCET}\,.
\end{aligned}
\end{eqnarray}
%%%%%%%%%%%%%%%%%%%%%%%%%%%%%%%%%%%%%%%%%%%%%%%%%
Comparing with the experimental data Eq.~(\ref{expR}), we can see that (i) within the Scenario S4 of QCDF and the SCET formalism, even though the ratio $R_{+-}^{\pi\pi}$ is well predicted, their predictions for $R_{00}^{\pi\pi}$ are still much lower than the data; (ii) the pQCD predictions are, on the other hand, even worse no matter the NLO corrections are included or not. It is also noted that, although a large electro-weak penguin~(EWP) contribution could resolve the anomaly observed in $B\to\pi\pi$ and $\pi K$ decays~\cite{pipipuz1,pipipuz2,PiPiEWP1,PiPiEWP2}, there are no known mechanisms to enhance such a large EWP amplitude. Furthermore, a large color-suppressed tree amplitude, which is also helpful to moderate these puzzles~\cite{PiPipQCD,PiPiCS1,PiPiCS2,PiPiCS3,PiPiCS4}, is hard to obtain from any short-distance dynamics. Thus, it is quite
difficult to understand these anomalies within the SM. Many efforts have been made to bridge these large discrepancies both within the SM~\cite{PiPiSM1,PiPiSM2,PiPiSM3,PiPiSM4,PiPiSM5,PiPiSM6,PiPiSM7,PiPiSM8,PiPiSM9} and in various new physics~(NP) scenarios~\cite{PiPiNP1,PiPiNP2,PiPiNP3,PiPiNP4,PiPiNP5}.

As is well-known, an additional $U(1)^{\prime}$ gauge symmetry and associated $Z^{\prime}$ gauge boson could arise in some well-motivated extensions of the SM. Searching for such an extra $Z^{\prime}$ boson is an important mission in the experimental programs of Tevatron~\cite{Tevatron} and LHC~\cite{LHC1,LHC2}. Performing
the constraints on the new $Z^{\prime}$ couplings through low-energy physics is, on the other hand, very
crucial and complementary for these direct searches. Theoretically, one of the simple extensions is the family non-universal $Z^{\prime}$ model, which could be naturally derived in certain string constructions~\cite{string1,string2,string3}, $E_6$ models~\cite{E61,E62,E63,E64,E65} and so on. It is interesting to note that the non-universal $Z^{\prime}$ couplings could lead to large flavour-changing neutral current~(FCNC) processes
and possible new source of CP-violating effects~\cite{Langacker1,Langacker2}. Such a specific model could reconcile the anomalies observed in $B_s-\bar{B}_s$ mixing and $B\to\phi K_s$ decay~\cite{Liu1,Liu2,Barger1,Barger2},
the ``$\pi K$ puzzle"~\cite{Barger,Chang,Giri}, and the anomalous polarization in $B\to\phi K^{\ast}$
decays~\cite{Chen:2006vs}. It has also been examined in some other interesting processes~\cite{Chang:2010zy1,Chang:2010zy2,Hua:2010wf1,Hua:2010wf2,Hua:2010wf3,Hua:2010wf4,Chiang:2006we,BZprime1,BZprime2}.

Based on the above observations, in this paper we pursue a possible solution to the observed ``$\pi\pi$'' puzzle within such a family non-universal $Z^{\prime}$ model. We shall adopt the QCDF approach~\cite{Beneke11,Beneke12,Beneke13,Beneke3} to evaluate the relevant hadronic matrix elements of $B \to \pi\pi$ decays, with an alternative scheme to parameterize the end-point divergence appearing in hard-spectator and annihilation corrections~\cite{Chang,YDYang}. In addition, since only the flavour-nondiagonal couplings $B_{pb}$~($p=d,s$) are different between the quark-level transitions $b\to d\bar{q}q$ and $b\to s\bar{q}q$~(with $q=u$ or $d$) within the model, it is interesting to investigate if the allowed parameter spaces in $B\to\pi\pi$ decays could survive after taking into account the constraints from $B\to\pi K, \pi K^{\ast}$ and $\rho K$ decays~\cite{Chang}.

Our paper is organized as follows. In Section~2, a survey of $B\to\pi \pi$ decays in the SM within the QCDF formalism is given; our numerical result, with two different schemes for the end-point divergence, is also presented. In Section~3, after briefly reviewing the non-universal $Z^{\prime}$ model, we present our numerical results and discussions in detail. Section~4 contains our conclusions. Appendix~A recapitulates the SM decay amplitudes of the $B\to\pi \pi$ decays within the QCDF formalism~\cite{Beneke11,Beneke12,Beneke13,Beneke3}. Appendix~B contains the relevant formulas for  hard-spectator and annihilation amplitudes with the infrared finite gluon propagator~\cite{Chang,YDYang}. All the theoretical input parameters are summarized in Appendix~C.

\section{Revisiting $B\to\pi\pi$ decays within the QCDF framework}
\label{sec:SM}

In the SM, the effective weak Hamiltonian responsible for $b\to d$ transition is given
as~\cite{Buchalla:1996vs1,Buchalla:1996vs2}
%%%%%%%%%%%%%%%%%%%%%%%%%%%%%%%%%%%%%%%%%%%%%%%%%
\begin{eqnarray}\label{eq:eff}
 {\cal H}^{\rm SM}_{\rm eff} &=& \frac{G_F}{\sqrt{2}} \biggl[V_{ub}
 V_{ud}^* \left(C_1 O_1^u + C_2 O_2^u \right) + V_{cb} V_{cd}^* \left(C_1
 O_1^c + C_2 O_2^c \right) - V_{tb} V_{td}^*\, \big(\sum_{i = 3}^{10}
 C_i O_i \big. \biggl. \nonumber\\
 && \biggl. \big. + C_{7\gamma} O_{7\gamma} + C_{8g} O_{8g}\big)\biggl] +
 {\rm h.c.},
\end{eqnarray}
%%%%%%%%%%%%%%%%%%%%%%%%%%%%%%%%%%%%%%%%%%%%%%%%%
where $V_{qb} V_{qd}^*$~($q=u$, $c$ and $t$) are products of the Cabibbo-Kobayashi-Maskawa~(CKM) matrix elements~\cite{ckm1,ckm2}, $C_{i}$ the Wilson coefficients, and $O_i$ the relevant four-quark operators whose explicit forms could be found, for example, in Refs.~\cite{Beneke11,Beneke12,Beneke13,Buchalla:1996vs1,Buchalla:1996vs2}.

In recent years, the QCDF approach has been employed extensively to study hadronic B-meson decays. For example, the tree-dominated $B\to\pi\pi$ decays have been studied comprehensively within the SM in Refs.~\cite{Beneke11,Beneke12,Beneke13,Beneke3}. It is also noted that the framework contains estimates of the hard-spectator and annihilation corrections. Even though they are power-suppressed, their strengthes and associated strong-interaction phases are numerically important to reproduce the experimental data~\cite{Beneke3,PiPiCS1,PiPiCS2,PiPiCS3,PiPiCS4}. Unfortunately, there are end-point divergences associated with the twist-3 hard-spectator and the annihilation amplitudes. Thus, how to regulate the divergence becomes indispensable within the QCDF formalism.

%%%%%%%%%%%%%Table 1%%%%%%%%%%%%%%%%%%%%%%%%%%%
\begin{table}[t]
 \begin{center}
 \caption{\small The $CP$-averaged branching ratios~(in units of $10^{-6}$) and the CP asymmetries~(in unit of $10^{-2}$) of $B\to \pi\pi$ decays in the SM with the two regulation schemes for the end-point divergence.}
 \label{tabpipiSM}
 \vspace{0.1cm}
 \doublerulesep 0.7pt \tabcolsep 0.04in
 \begin{tabular}{lccccccc} \hline \hline
 \multicolumn{1}{c}{Observable}             &\multicolumn{1}{c}{Exp. data}&\multicolumn{1}{c}{Scheme I}  &\multicolumn{3}{c}{Scheme II}\\
                                            &                      &              &$m_g=0.3~{\rm GeV}$ &$m_g=0.7~{\rm GeV}$ &$m_g=0.5\pm0.05~{\rm GeV}$     \\
 \hline
 ${\cal B}(B^-\to\pi^-\pi^0)$             &$5.59^{+0.41}_{-0.40}$  &$4.64\pm0.74$   &$5.42\pm0.65$  &$4.01\pm0.64$ &$4.52\pm0.70$ \\
 ${\cal B}({\bar{B}}^0\to\pi^+\pi^-)$     &$5.16\pm0.22$           &$6.88\pm1.13$   &$9.32\pm1.07$ &$7.44\pm1.00$ &$8.14\pm1.21$ \\
 ${\cal B}({\bar{B}}^0\to\pi^0\pi^0)$     &$1.55\pm0.19$           &$0.77\pm0.12$ &$2.03\pm0.24$ &$0.57\pm0.07$ &$0.95\pm0.18$ \\
 \hline
 $A_{CP}(B^-\to\pi^-\pi^0)$               &$6\pm5$                 &$-0.1\pm0.1$  &$-0.1\pm0.1$&$-0.1\pm0.1$    &$-0.1\pm0.1$  \\
 $A_{CP}({\bar{B}}^0\to\pi^+\pi^-)$       &$38\pm6$                &$15\pm2$      &$42\pm3$      &$18\pm2$      &$26\pm3$  \\
 $A_{CP}({\bar{B}}^0\to\pi^0\pi^0)$       &$43^{+25}_{-24}$        &$-7\pm6$       &$-60\pm4$     &$-35\pm4$     &$-45\pm5$  \\
 \hline
 $A_{CP}^{mix}({\bar{B}}^0\to\pi^+\pi^-)$ &$-65^{+7}_{-7}$         &$-58\pm7$     &$-69\pm6$     &$-52\pm6$     &$-64\pm7$  \\
 $A_{CP}^{mix}({\bar{B}}^0\to\pi^0\pi^0)$ &---                     &$70\pm10$     &$17\pm4$      &$45\pm4$      &$33\pm6$   \\
 \hline\hline
 \end{tabular}
 \end{center}
 \end{table}
%%%%%%%%%%%%%%%%%%%%%%%%%%%%%%%%%%%%%%%%%%%%

As the usually adopted scheme~(Scheme~I), the divergent integrals are phenomenologically parameterized by
$X_{A}$~(annihilation) and $X_{H}$~(hard spectator scattering)~\cite{Beneke11,Beneke12,Beneke13,Beneke3}, with
%%%%%%%%%%%%%%%%%%%%%%%%%%%%%%%%%%%%%%%%%%%%%%%%%
\begin{equation}\label{treat-for-anni}
\int_0^1 \frac{\!dy}{y}\, \to X_A =(1+\rho_A e^{i\phi_A}) \ln
\frac{m_B}{\Lambda_h}, \qquad \int_0^1dy \frac{\textmd{ln}y}{y}\,
\to -\frac{1}{2}(X_A)^2\,,
 \end{equation}
%%%%%%%%%%%%%%%%%%%%%%%%%%%%%%%%%%%%%%%%%%%%%%%%%
and similarly for $X_{H}$. Here $\Lambda_h=0.5\,{\rm GeV}$, $\rho_{H,A} \leq 1$ and $\phi_{H,A}$ is an arbitrary strong-interaction phase, which might be caused by soft rescattering. The different choices of $\rho_A$ and $\phi_A$ correspond to the different scenarios discussed in Ref.~\cite{Beneke3}. In our numerical evaluations, we take $\rho_{H,A}=1$ and $\phi_{H,A}=-55^{\circ}$, which are suggested by the most favorable scenario S4 in Ref.~\cite{Beneke3}. For estimating theoretical uncertainties, we shall assign an error of $\pm0.1$ to $\rho_{H,A}$ and $\pm5^{\circ}$ to $\phi_{H,A}$.

As the above parametrization is rather arbitrary, it is still very worthy to find some alternative schemes to regulate these endpoint divergences, as precisely as possible, to estimate the strengths and the associated strong phases in these power-suppressed contributions. As an alternative scheme~(Scheme~II), we shall use the infrared finite gluon propagator to regulate the end-point divergence. It is interesting to note that recent theoretical and phenomenological studies are now accumulating supports for a softer infrared behavior of the gluon propagator~\cite{Alkofer1,Alkofer2,Alkofer3,Alkofer4,theo1,theo2,theo3,theo4,phe1,phe2,phe3}, which is also indicated by recent lattice simulations~\cite{Cucchieri:2007md1,Cucchieri:2007md2}. Furthermore, the infrared finite dynamical gluon propagator, which is shown to be not divergent as fast as $1/q^2$, has been applied successfully to two-body hadronic B-meson decays~\cite{Chang,YDYang,YYgluon1,YYgluon2,YYgluon3,Natale,Zanetti}. To be specific, in our calculation we shall adopt the gluon propagator derived by Cornwall~(in Minkowski space)~\cite{Cornwall1,Cornwall2}
%%%%%%%%%%%%%%%%%%%%%%%%%%%%%%%%%%%%%%%%%%%%%%%%%
\begin{eqnarray}
D(q^2)=\frac{1}{q^2-M_g^2(q^2)+i\epsilon}~,
 \label{Dg}
\end{eqnarray}
%%%%%%%%%%%%%%%%%%%%%%%%%%%%%%%%%%%%%%%%%%%%%%%%%
where the dynamical gluon mass square $M_g^2(q^2)$ is obtained as~\cite{Cornwall1,Cornwall2}
%%%%%%%%%%%%%%%%%%%%%%%%%%%%%%%%%%%%%%%%%%%%%%%%%
\begin{eqnarray}
M_g^2(q^2)=m_g^2\Bigg[\frac{\mathrm{ln}\Big(\frac{q^2+4m_g^2}{\Lambda_{\rm
QCD}^2}\Big)} {\mathrm{ln}\Big(\frac{4m_g^2}{\Lambda_{\rm
QCD}^2}\Big)}\Bigg]^{-\frac{12}{11}}, \label{Mg}
\end{eqnarray}
%%%%%%%%%%%%%%%%%%%%%%%%%%%%%%%%%%%%%%%%%%%%%%%%%
with $q$ being the gluon momentum, $m_g$ the effective gluon mass and $\Lambda_{\rm QCD}=225~{\rm MeV}$. The corresponding strong coupling constant reads
%%%%%%%%%%%%%%%%%%%%%%%%%%%%%%%%%%%%%%%%%%%%%%%
\begin{eqnarray}
\alpha_s(q^2)=\frac{4\pi}{\beta_0\mathrm{ln}\Big(\frac{q^2+4M_g^2(q^2)}
{\Lambda_{QCD}^2}\Big)}~, \label{Alphas}
\end{eqnarray}
%%%%%%%%%%%%%%%%%%%%%%%%%%%%%%%%%%%%%%%%%%%%%%%%%
where $\beta_0=11-\frac{2}{3}n_f$ is the first coefficient of the QCD beta function, and $n_f$ the number of active quark flavors.

With such a scheme, the hard-spectator scattering contributions are real, while the annihilation contributions are complex with a large imaginary part~\cite{YDYang}. Both the hard-spectator scattering and the annihilation contributions are very sensitive to the value of the effective gluon mass~\cite{YDYang,Zanetti}, with a typical value $m_g=500\pm200~{\rm MeV}$~\cite{Cornwall1,Cornwall2}. In Ref.~\cite{YDYang}, we presented our suggested value, $m_g=500\pm50~{\rm MeV}$, which is a reasonable choice so that most of the observables for $B\to\pi K$,~$\pi K^{\ast}$ and $\rho K$ decays are in good agreement with the experimental data. Furthermore, comparing to the available data on $B^0\to K^+K^-$,$D_s^{(\ast)}K$, $B_s\to\pi\pi$ and so on, Natale and Zanetti also found that the gluon mass scale is close to $500~{\rm MeV}$~\cite{Zanetti}. So, in this paper, we shall take $500~{\rm MeV}$ as the central value of the gluon mass scale and $\pm 50~{\rm MeV}$ as its uncertainty. As a comparison, the numerical results with $m_g=700~{\rm MeV}$ and $300~{\rm MeV}$ are also presented in Table~\ref{tabpipiSM}.

With the above two schemes, our SM predictions for the branching ratios, the direct and the mixing-induced CP asymmetries of $B\to\pi\pi$ decays are listed in Table~\ref{tabpipiSM}. We find that, although ${\cal B}({\bar{B}}^0\to\pi^0\pi^0)$ in Scheme~II could be enhanced relative to the one based on Scheme~I, ${\cal B}({\bar{B}}^0\to\pi^+\pi^-)$ is also increased at the same time. Thus, as expected within the SM, we get the ratios $R_{+-}^{\pi\pi}=1.04\pm0.22$ and $R_{00}^{\pi\pi}=0.23\pm0.06$, which are $3.5\sigma$ and $3.7\sigma$ lower than the current experimental data Eq.~(\ref{expR}), respectively. In addition, our prediction $A_{CP}({\bar{B}}^0\to\pi^0\pi^0)\sim-0.45$, although being roughly consistent with the central value in Scheme~I $\sim-0.07$ and that in SCET $\sim-0.58$~\cite{PiPiSCET}, is still quite different from the current data $0.43_{-0.24}^{+0.25}$~\cite{HFAG}. In the following, we shall investigate if these mismatches could be reconciled within a family non-universal $Z^{\prime}$ model.

\section{Solution within the family non-universal $Z^{\prime}$ model}
\label{sec:Z-prime}

\subsection{Formalism of the family non-universal $Z^{\prime}$ model}

A family non-universal $Z^{\prime}$ model can lead to FCNC processes even at tree level due to the non-diagonal chiral coupling matrix. The formalism of the model has been detailed in Ref.~\cite{Langacker1,Langacker2}.
The relevant studies in the context of B physics have also been extensively performed in Refs.~\cite{Liu1,Liu2,Barger1,Barger2,Barger,Chang,Giri,Chen:2006vs,Chang:2010zy1,Chang:2010zy2,Hua:2010wf1,Hua:2010wf2,Hua:2010wf3,Hua:2010wf4,BZprime1,BZprime2}.

With the assumption of flavor-diagonal right-handed couplings and neglecting the kinetic mixing term as
usually adopted in the literature, the $Z^{\prime}$ part of the effective Hamiltonian for $b\to p\bar{q}q$~($p=d,s$ and $q$ denotes the active quarks) transitions can be written
as~\cite{Langacker1,Langacker2,Liu1,Liu2,Barger}
%%%%%%%%%%%%%%%%%%%%%%%%%%%%%%%%%%%%%%%%%%%%%%%%%
\begin{equation}\label{heffz1}
 {\cal H}_{\rm eff}^{\rm
 Z^{\prime}}=\frac{2G_F}{\sqrt{2}}\big(\frac{g^{\prime}M_Z}
 {g_1M_{Z^{\prime}}}\big)^2
 \,B_{pb}^L(\bar{p}b)_{V-A}\sum_{q}\big[B_{qq}^L (\bar{q}q)_{V-A}
 +B_{qq}^R(\bar{q}q)_{V+A}\big]+{\rm h.c.}\,,
\end{equation}
%%%%%%%%%%%%%%%%%%%%%%%%%%%%%%%%%%%%%%%%%%%%%%%%%
where $g_1=e/(\sin{\theta_W}\cos{\theta_W})$, $g^{\prime}$ is the gauge coupling constant of extra
$U^{\prime}(1)$ group, $M_{Z^{\prime}}$ the mass of the new gauge boson, and $B_{ij}^X$ refer to the effective $Z^{\prime}$ couplings to the quarks $i$ and $j$ at the electro-weak scale. It is noted that the forms of four-quark operators in Eq.~(\ref{heffz1}) already exist in the SM. As a result, we can rewrite
Eq.~(\ref{heffz1}) as
%%%%%%%%%%%%%%%%%%%%%%%%%%%%%%%%%%%%%%%%%%%%%%%%%
\begin{equation}
 {\cal H}_{\rm eff}^{\rm
 Z^{\prime}}=-\frac{G_F}{\sqrt{2}}V_{tb}V_{tp}^{\ast}\sum_{q}
 \big(\Delta C_3 O_3^q +\Delta C_5 O_5^q+\Delta C_7 O_7^q+\Delta C_9
  O_9^q\big)+{\rm h.c.}\,,
\end{equation}
%%%%%%%%%%%%%%%%%%%%%%%%%%%%%%%%%%%%%%%%%%%%%%%%%
where $O_i^q~(i=3,5,7,9)$ are the effective four-quark operators in the SM, and $\Delta C_i$
denote the modifications to the corresponding SM Wilson coefficients induced by the new gauge boson
$Z^{\prime}$, which are expressed as
%%%%%%%%%%%%%%%%%%%%%%%%%%%%%%%%%%%%%%%%%%%%%%%%%
\begin{eqnarray}
 \Delta C_{3,5}&=&-\frac{2}{3V_{tb}V_{tp}^{\ast}}\,\big(\frac{g^{\prime}M_Z}
 {g_1M_{Z^{\prime}}}\big)^2\,B_{pb}^L\,(B_{uu}^{L,R}+2B_{dd}^{L,R})\,,\nonumber\\
 \Delta C_{9,7}&=&-\frac{4}{3V_{tb}V_{tp}^{\ast}}\,\big(\frac{g^{\prime}M_Z}
 {g_1M_{Z^{\prime}}}\big)^2\,B_{pb}^L\,(B_{uu}^{L,R}-B_{dd}^{L,R})\,,
 \label{NPWilson}
\end{eqnarray}
%%%%%%%%%%%%%%%%%%%%%%%%%%%%%%%%%%%%%%%%%%%%%%%%%
in terms of the model parameters at the initial scale $\mu_{W}\sim m_W$~(the $W^{\pm}$ boson mass).

Generally, the diagonal elements of the effective coupling matrices $B_{qq}^{L,R}$ are real as a consequence of the hermiticity of the effective weak Hamiltonian. However, the off-diagonal ones $B_{pb}^L$ can contain a new weak phase $\phi^L_p$. Then, for convenience we can represent $\Delta C_i$ as
%%%%%%%%%%%%%%%%%%%%%%%%%%%%%%%%%%%%%%%%%%%%%%%%%
\begin{eqnarray}\label{deltac3597}
 \Delta
 C_{3,5}=2\,\frac{|V_{tb}V_{tp}^{\ast}|}{V_{tb}V_{tp}^{\ast}}\,
 \zeta^{LL,LR}_p\,e^{i\phi^L_p}\,,\quad
 \Delta
 C_{9,7}=4\,\frac{|V_{tb}V_{tp}^{\ast}|}{V_{tb}V_{tp}^{\ast}}\,
 \xi^{LL,LR}_p\,e^{i\phi^L_p}\,,
\end{eqnarray}
%%%%%%%%%%%%%%%%%%%%%%%%%%%%%%%%%%%%%%%%%%%%%%%%%
where the newly introduced $Z^{\prime}$ parameters $\zeta^{LL,LR}_p$, $\xi^{LL,LR}_p$ and $\phi^L_p$ are defined, respectively, as
%%%%%%%%%%%%%%%%%%%%%%%%%%%%%%%%%%%%%%%%%%%%%%%%%
\begin{eqnarray}\label{zetaandxi}
 \zeta^{LL,LR}_p&=&-\frac{1}{3}\,\big(\frac{g^{\prime}M_Z}
 {g_1M_{Z^{\prime}}}\big)^2\,\big|\frac{B_{pb}^L}{V_{tb}V_{tp}^{\ast}}\big|\,
 (B_{uu}^{L,R}+2B_{dd}^{L,R})\,,\nonumber\\[0.2cm]
 \xi^{LL,LR}_p&=&-\frac{1}{3}\,\big(\frac{g^{\prime}M_Z}{g_1M_{Z^{\prime}}}\big)^2\,
 \big|\frac{B_{pb}^L}{V_{tb}V_{tp}^{\ast}}\big|\,(B_{uu}^{L,R}-B_{dd}^{L,R})\,,
 \nonumber\\[0.2cm]
 \phi^L_p&=&\arg{[B_{pb}^L]}\,.
\end{eqnarray}
%%%%%%%%%%%%%%%%%%%%%%%%%%%%%%%%%%%%%%%%%%%%%%%%%
It is noted that the other SM Wilson coefficients may also receive contributions from the $Z^{\prime}$ boson through renormalization group~(RG) evolution. With our assumption of no significant RG running effect from the scales $M_Z^{\prime}$ to $M_W$, the RG evolution of the modified Wilson coefficients is exactly the same as that in the SM~\cite{Buchalla:1996vs1,Buchalla:1996vs2}. Numerical results of the Wilson coefficients at the lower scales $m_b$ and $\sqrt{\Lambda_h m_b}$ are listed in Appendix~C1.

In our analyses, we also define the following three ratios~(taking $\xi^{LL}_p$ as a benchmark)
%%%%%%%%%%%%%%%%%%%%%%%%%%%%%%%%%%%%%%%%%%%%%%%%%
\begin{eqnarray}\label{Ratio}
 R_1\equiv\frac{\xi^{LR}_p}{\xi^{LL}_p}=\frac{B_{uu}^{R}-B_{dd}^{R}}{B_{uu}^{L}-B_{dd}^{L}}\,,\quad
 R_2\equiv\frac{\zeta^{LL}_p}{\xi^{LL}_p}=\frac{B_{uu}^{L}+2B_{dd}^{L}}{B_{uu}^{L}-B_{dd}^{L}}\,,\quad
 R_3\equiv\frac{\zeta^{LR}_p}{\xi^{LL}_p}=\frac{B_{uu}^{R}+2B_{dd}^{R}}{B_{uu}^{L}-B_{dd}^{L}}\,.
\end{eqnarray}
%%%%%%%%%%%%%%%%%%%%%%%%%%%%%%%%%%%%%%%%%%%%%%%%%
It is interesting to note that these ratios are independent of the quark flavor $p\,(p=d,s)$, {\it i.e.}, they are the same for both $b\to s \bar{q}q$ and $b\to d \bar{q}q$ transitions. In fact, there are only two different $Z^{\prime}$ parameters $|B_{pb}|$ and $\phi^L_p$ between these two types of transitions. Thus, the $Z^{\prime}$ parameter spaces constrained by $B\to\pi\pi$ decays should also suffer constraints from $B\to\pi K$, $\pi K^{\ast}$ and $\rho K$ decays, which have been investigated in detail in our previous paper~\cite{Chang}.

\subsection{Numerical results and discussions}

With the theoretical formulae and the input parameters summarized in Appendices A, B and C, we now present our numerical analyses and discussions. We take the current experimental data on ${\cal B}(B\to\pi\pi)$ and $A_{CP}(B\to\pi\pi)$ as constraints and $A_{CP}^{mix}(B\to\pi^{+}\pi^{-}, \pi^{0}\pi^{0})$ as our theoretical predictions. Our fitting is performed with the experimental data varying randomly within $2\sigma$ error-bars, while the theoretical uncertainties are obtained by varying the input parameters within the regions specified in Appendix C. For simplicity, in the analysis and discussion of this subsection, we mainly adopt the Scheme~II with $m_g=0.5\pm0.05\,{\rm GeV}$ to regulate the end-point divergence. While, as a comparison, the fitting results with scheme~I are also presented.

\begin{figure}[htbp]
\begin{center}
\epsfxsize=15cm \centerline{\epsffile{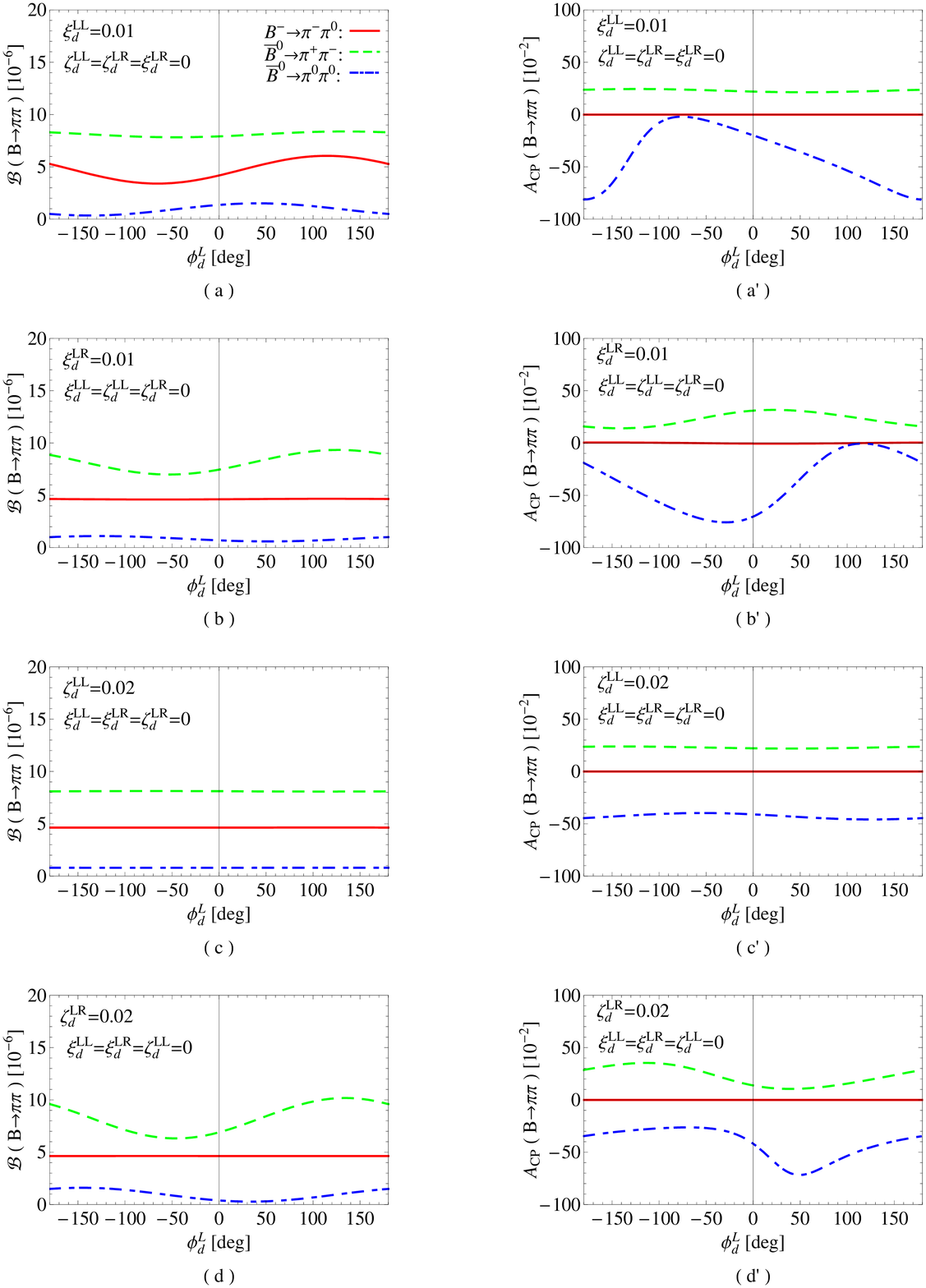}}
\centerline{\parbox{16cm}{\caption{\label{CPBPhiL}\small (Color online) The dependence of ${\cal B}(B\to\pi\pi)$ and $A_{CP}(B\to\pi\pi)$ on the new weak phase $\phi_d^L$ with the values of $\xi^{LL}_d$, $\xi^{LR}_d$, $\zeta^{LL}_d$ and $\zeta^{LR}_d$ marked by the legends. }}}
\end{center}
\end{figure}

The dependence of the observables ${\cal B}(B\to\pi\pi)$ and $A_{CP}(B\to\pi\pi)$ on the new weak phase $\phi^L_d$, with definite values of $\xi^{LL}_d$, $\xi^{LR}_d$, $\zeta^{LL}_d$ and $\zeta^{LR}_d$ marked in the legends, is shown in Fig.~\ref{CPBPhiL}. We can see that both ${\cal B}(B\to\pi\pi)$ and $A_{CP}(B\to\pi\pi)$ are most sensitive to the $Z^{\prime}$ contributions involving $\xi^{LL}_d$, $\xi^{LR}_d$ and $\zeta^{LR}_d$, but not to the ones involving $\zeta^{LL}_d$. Thus, as shown in Figs.~\ref{CPBPhiL}(c) and ($\rm{c}^{\prime}$), the $Z^{\prime}$ effects involving $\zeta^{LL}_d$ could always be neglected compared with the other ones. To be more specific, Our analyses are divided into the following three different cases:
%%%%%%%%%%%%%%%%%%%%%%%%%%%%%%%%%%%%%
\begin{itemize}
\item Case I: With only $\zeta^{LR}_d$ arbitrary, and neglecting all the others;

\item Case II: With $\zeta^{LR}_d$ and $\xi^{LL,LR}_d$ arbitrary, but neglecting $\zeta^{LL}_d$;

\item Case III: Combining with the fitting results from $B\to\pi K$, $\pi K^{\ast}$ and $\rho K$ decays.
\end{itemize}
%%%%%%%%%%%%%%%%%%%%%%%%%%%%%%%%%%%%%
Corresponding to such three cases, our fitting results for the $Z^{\prime}$ parameters are summarized in Table~\ref{NPPareValue}. With these values of $Z^{\prime}$ parameters as input, our predictions for the observables in $B\to\pi\pi$ decays are listed in Table~\ref{tabpipiZp}.

%%%%%%%%%%%%%%%%%%%%%%%%%%%%%%%%%%%%%%%%
\begin{table}
 \begin{center}
 \caption{\small Numerical results for the $Z^{\prime}$ coupling parameters $\xi^{LL,LR}_d$, $\zeta^{LR}_d$ and $\phi^L_d$ in the three different specific cases with the regulation scheme~II~(scheme~I) for the end-point divergence.  The dashes in Case~I means that the corresponding parameters are neglected.}
 \label{NPPareValue}
 \vspace{0.1cm}
 \doublerulesep 0.7pt \tabcolsep 0.10in
 \begin{tabular}{lccc} \hline \hline
  Parameters                    &Case I                     &Case II                       &Case III     \\\hline
 $\xi^{LL}_d(\times 10^{-3})$   &---                        &$-5.8\pm6.3$~$(-5.9\pm4.9)$   &$7.6\pm1.2$  \\
 $\xi^{LR}_d(\times 10^{-3})$   &---                        &$-13.2\pm7.5$~$(-13.5\pm8.6)$ &$-2.4\pm1.7$ \\
 $\zeta^{LR}_d(\times 10^{-3})$ &$48.5\pm5.2$~$(37.0\pm5.4)$&$49.5\pm6.9$~$(41.8\pm9.0)$   &$45.4\pm5.0$  \\
 $\phi^{L}_d(^{\circ})$         &$-51\pm5$~$(-71\pm6)$      &$-52\pm6$~$(-68\pm7)$         &$-48\pm6$ \\
 \hline \hline
 \end{tabular}
 \end{center}
 \end{table}
%%%%%%%%%%%%%%%%%%%%%%%%%%%%%%%%%%%%%%%%%%%%

%%%%%%%%%%%%%Table 2%%%%%%%%%%%%%%%%%%%%%%%%%%%
\begin{table}[t]
 \begin{center}
 \caption{\small The $CP$-averaged branching ratios~(in units of $10^{-6}$) and the CP asymmetries~(in unit of $10^{-2}$) of $B\to \pi\pi$ decays within the non-universal $Z^{\prime}$ model with the regulation scheme~II for the end-point divergence. The theoretical errors of our predictions in Case~I,~II and III correspond (in this order) to the uncertainties of ``parameters listed in appendix~C and $m_g$'' and ``$Z^{\prime}$ parameters''.}
 \label{tabpipiZp}
 \vspace{0.1cm}
 \doublerulesep 0.7pt \tabcolsep 0.04in
 \begin{tabular}{lccccc} \hline \hline
 \multicolumn{1}{c}{Observable}          &\multicolumn{1}{c}{Exp. data}&\multicolumn{1}{c}{SM}&\multicolumn{3}{c}{$Z^{\prime}$ model} \\
                                         &                                 &    &Case I               &Case II       &Case III \\
 \hline
 ${\cal B}(B^-\to\pi^-\pi^0)$            &$5.59^{+0.41}_{-0.40}$ &$4.52\pm0.70$ &$4.64\pm0.61\pm0.0$    &$5.46\pm0.72\pm0.83$ &$4.10\pm0.51\pm0.12$  \\
 ${\cal B}({\bar{B}}^0\to\pi^+\pi^-)$    &$5.16\pm0.22$          &$8.14\pm1.21$ &$4.52\pm0.68\pm0.23$ &$5.71\pm0.81\pm0.93$ &$4.66\pm0.64\pm0.29$ \\
 ${\cal B}({\bar{B}}^0\to\pi^0\pi^0)$    &$1.55\pm0.19$          &$0.95\pm0.18$ &$1.41\pm0.20\pm0.18$ &$1.60\pm0.23\pm0.32$ &$1.16\pm0.16\pm0.14$\\
 \hline
 $A_{CP}(B^-\to\pi^-\pi^0)$              &$6\pm5$                &$-0.1\pm0.1$ &$-0.1\pm0.1\pm0.0$     &$0.1\pm0.1\pm0.8$    &$0.0\pm0.0\pm0.3$ \\
 $A_{CP}({\bar{B}}^0\to\pi^+\pi^-)$      &$38\pm6$               &$26\pm3$ &$32\pm5\pm5$         &$29\pm4\pm6$         &$30\pm5\pm5$     \\
 $A_{CP}({\bar{B}}^0\to\pi^0\pi^0)$      &$43^{+25}_{-24}$       &$-45\pm5$ &$-8\pm4\pm2$         &$11\pm3\pm7$        &$9\pm3\pm4$   \\
 \hline
 $A_{CP}^{mix}({\bar{B}}^0\to\pi^+\pi^-)$&$-65\pm7$              &$-64\pm7$ &$-53\pm8\pm4$        &$-52\pm7\pm4$        &$-50\pm8\pm4$  \\
 $A_{CP}^{mix}({\bar{B}}^0\to\pi^0\pi^0)$&---                    &$33\pm6$ &$82\pm8\pm7$         &$83\pm8\pm13$        &$90\pm5\pm4$    \\
 \hline\hline
 \end{tabular}
 \end{center}
 \end{table}
%%%%%%%%%%%%%%%%%%%%%%%%%%%%%%%%%%%%%%%%%%%%

%%%%%%%%%%%%%%%%%%%%%%%%%%%%%%%%%%%%%%%%
%%%%%%%%%%%%%%%%%%%%%%%%%%%%%%%%%%%%%%%%%%%

\subsubsection*{Case~I: With only $\zeta^{LR}_d$ arbitrary, and neglecting all the others.}

In this specific case, as can be seen from Eqs.~(\ref{deltac3597}) and (\ref{zetaandxi}), the $Z^{\prime}$ contribution is mainly embodied in the QCD penguin sector $\Delta C_5$. From Figs.~\ref{CPBPhiL}(b) and (d), with $\xi^{LR}_d\sim\mathcal{O}(10^{-2})$ and/or $\zeta^{LR}_d\sim\mathcal{O}(10^{-2})$ respectively, we can see that ${\cal B}(\bar{B}^0\to\pi^+\pi^-)$ is reduced to be consistent with the experimental data at $\phi^L_d\sim-50^{\circ}$, which is helpful to reconcile the ``$\pi\pi$ puzzle''. However, as shown in Fig.~\ref{CPBPhiL}($\rm{b}^{\prime}$), the $Z^{\prime}$ contributions involving a positive $\xi^{LR}_d$ at
$\phi^L_d\sim-50^{\circ}$ induce a large negative $A_{CP}(\bar{B}^0\to\pi^0\pi^0)$, which is in obvious conflict with the experimental measurement $0.43^{+0.25}_{-0.24}$. As a consequence, the case with $Z^{\prime}$ contributions involving a positive $\xi^{LR}_d$ at $\phi^L_d\sim-50^{\circ}$ could be excluded, and we are left with a possible solution to the ``$\pi\pi$ puzzle'' with only $\zeta^{LR}_d$ arbitrary as defined by Case~I.

\begin{figure}[t]
\begin{center}
\epsfxsize=7.5cm \centerline{\epsffile{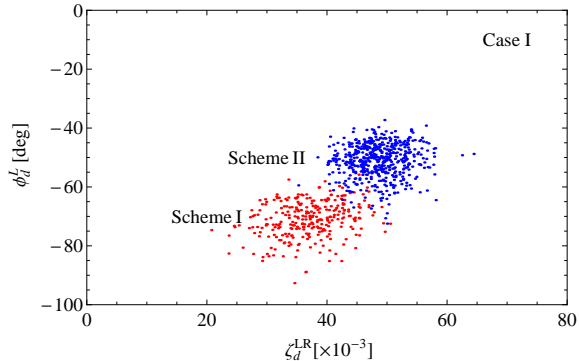}}
\centerline{\parbox{16cm}{\caption{\label{CaseI}\small (Color online) The final allowed regions for the $Z^{\prime}$ coupling parameters $\zeta^{LR}_d$ and $\phi^L_d$ in Case~I, constrained by the current experimental data on ${\cal B}(B\to\pi\pi)$ and $A_{CP}(B\to\pi\pi)$.}}}
\end{center}
\end{figure}

Corresponding to regulation scheme I and scheme II for the end-point divergency respectively, the final allowed regions for the $Z^{\prime}$ coupling parameters $\zeta^{LR}_d$ and $\phi^L_d$ under the constraints from ${\cal B}(B\to\pi\pi)$ and $A_{CP}(B\to\pi\pi)$ are shown in Fig.~\ref{CaseI}. The corresponding numerical results are listed in the second column of Table~\ref{NPPareValue}. Obviously, one may find that the result of $\zeta^{LR}_d$~($|\phi^L_d|$) in scheme II is slightly larger~(smaller) than the one in scheme I.

From the fourth column of Table~\ref{tabpipiZp}, we can see that most of the observables in $B\to \pi \pi$ decays are consistent with the experimental measurements within $2\sigma$ error-bars. Most importantly, our theoretical predictions $R_{+-}^{\pi\pi}=1.92$ and $R_{00}^{\pi\pi}=0.63$ agree well with the experimental results $R_{+-}^{\pi\pi}=2.02\pm0.17$ and
$R_{00}^{\pi\pi}=0.60\pm0.08$~\cite{HFAG}, respectively, implying that the $Z^{\prime}$ contributions with $\zeta^{LR}_d\sim0.05$ and $\phi^{L}_d\sim-50^{\circ}$ are helpful to resolve the observed ``$\pi\pi$ puzzle''. It is also noted that, with such parameter values as inputs, our prediction for the mixing-induced CP asymmetry $A_{CP}^{mix}({\bar{B}}^0\to\pi^-\pi^+)$$\sim-0.53$, is in good agreement with the experimental measurement $-0.65\pm0.07$~\cite{HFAG}.

%%%%%%%%%%%%%%%%%%%%%%%%%%%%%%%%%%%%%%%%
%%%%%%%%%%%%%%%%%%%%%%%%%%%%%%%%%%%%%%%%%%%

In this simplified case, the result $A_{CP}({\bar{B}}^0\to\pi^0\pi^0)\sim-8\%$ is, however, still $\sim2\sigma$ lower than the experimental data $(43^{+25}_{-24})\%$. Even though the $Z^{\prime}$ contributions with a large $\zeta^{LR}_d$ are helpful to enhance $A_{CP}({\bar{B}}^0\to\pi^0\pi^0)$, such a large value of $\zeta^{LR}_d$ is already excluded by ${\cal B}$$({\bar{B}}^0\to\pi^+\pi^-)$ as shown in Fig.~\ref{CPBPhiL}(d).

\subsubsection*{Case~II: With $\zeta^{LR}_d$ and $\xi^{LL,LR}_d$ arbitrary, but neglecting $\zeta^{LL}_d$.}

As discussed in Case~I, although most of the observables in $B\to \pi\pi$ decays could be accommodated, it is difficult to reproduce the observed direct CP asymmetry $A_{CP}({\bar{B}}^0\to\pi^0\pi^0)$. However, as shown in Figs.~\ref{CPBPhiL}($\rm{a}^{\prime}$) and ($\rm{b}^{\prime}$)~\footnote{The plots in the range $0\leqslant\phi_d^L\leqslant 180^{\circ}$ could be treated as the ones in $-180^{\circ}\leqslant\phi_d^L\leqslant 0$ by shifting the overall signs of $\xi^{LL,LR}_d$ and $\zeta^{LL,LR}_d$.}, this discrepancy could be possibly compensated by the extra $Z^{\prime}$ contributions involving positive $\xi^{LL}_d$ and/or negative $\xi^{LR}_d$ with $\phi^{L}_d\sim-50^{\circ}$. This observation motivates us to consider the second case defined by Case~II, where the parameters $\zeta^{LR}_d$ and $\xi^{LL,LR}_d$ are involved at the same time.

\begin{figure}[t]
\begin{center}
\epsfxsize=15cm \centerline{\epsffile{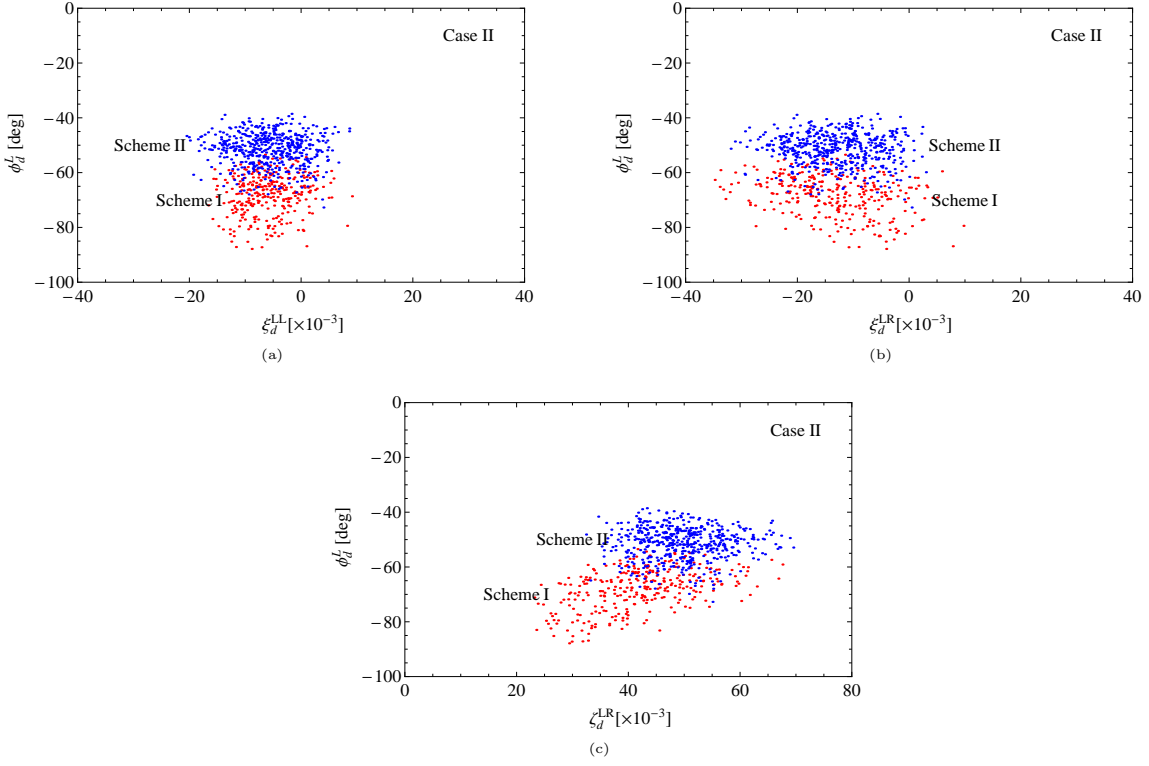}}
\centerline{\parbox{16cm}{\caption{\label{CaseII}\small (Color online) The final allowed regions for the $Z^{\prime}$ coupling parameters $\xi^{LL}_d$, $\xi^{LR}_d$, $\zeta^{LR}_d$ and $\phi^L_d$ in Case~II. The other captions are the same as in Fig.~\ref{CaseI}.}}}
\end{center}
\end{figure}

In this case, the final allowed regions for the four parameters $\xi^{LL}_d$, $\xi^{LR}_d$, $\zeta^{LR}_d$ and $\phi^L_d$ constrained by ${\cal B}(B\to\pi\pi)$ and $A_{CP}(B\to\pi\pi)$ are shown in  Fig.~\ref{CaseII}, and the corresponding numerical results are listed in the third column of Table~\ref{NPPareValue}. Similar to the situation in Case~I, we again find that the fitting value $|\phi^L_d|$ in scheme~I is a bit larger than the one in scheme~II. From Fig.~\ref{CaseII}, we can see that the value of $\zeta^{LR}_d$ is constrained to be definitely nonzero, which confirms our conclusion made in Case~I, {\it i.e.}, the $Z^{\prime}$ contributions with a NP weak phase $\phi^{L}_d\sim-50^{\circ}$ and $\zeta^{LR}_d\sim0.05$ are crucial to reconcile the observed ``$\pi\pi$ puzzle''. The constraints on $\xi^{LL}_d$ and $\xi^{LR}_d$ are, on the other hand, not very stringent and both of them could be even equal to zero, indicating that they play only a minor role in resolving the ``$\pi\pi$ puzzle''. However, our result $A_{CP}({\bar{B}}^0\to\pi^0\pi^0)=0.11\pm0.03\pm0.07$ shows that the effects involving $\xi^{LL,LR}_d$ are still very crucial to bridge the large discrepancy of $A_{CP}({\bar{B}}^0\to\pi^0\pi^0)$ encountered in Case~I.

%%%%%%%%%%%%%%%%%%%%%%%%%%%%%%%%%%%%%%%%
%%%%%%%%%%%%%%%%%%%%%%%%%%%%%%%%%%%%%%%%%%%

As discussed in detail in our previous publication~\cite{Chang}, the values of $\xi^{LL}_s$ and $\xi^{LR}_s$ are found to have different signs, which implies that the ratio $R_{1}$ defined in Eq.~(\ref{Ratio}) should be  negative. Since the ratio $R_{1}$ is independent of the quark flavor $d$ or $s$, the two parameters $\xi^{LL}_d$ and $\xi^{LR}_d$ are therefore also expected to have different signs. However, as shown in Figs.~\ref{CPBPhiL}(a) and \ref{CPBPhiL}$({\rm b}^{\prime})$, as well as in Fig.~\ref{CaseII}, a positive $\xi^{LL}_d$~($\xi^{LR}_d$) is suppressed~(even more strongly) by the observalbes ${\cal B}(B^-\to\pi^-\pi^0)$ and $A_{CP}(\bar{B}^0\to\pi^+\pi^-)$). Thus, after taking into account the fitting results of $R_{1,3}$ from Ref.~\cite{Chang}, the obtained ranges of $Z^{\prime}$ parameters $\zeta^{LL}_d$ and $\xi^{LL,LR}_d$ could be strongly reduced or even be excluded.

\subsubsection*{Case~III: Combining with the fitting results from $B\to\pi K$, $\pi K^{\ast}$ and $\rho K$ decays.}

In our previous work~\cite{Chang}, we have systemically investigated the effects of the family non-universal $Z^{\prime}$ model on penguin-dominated $B\to\pi K$, $\pi K^{\ast}$ and $\rho K$ decays, pursuing possible
solutions to the observed ``$\pi K$ puzzle''. We found that the model parameter spaces involving positive $\xi^{LL}_s$ and negative $\xi^{LR}_s$~(none of them could be neglected) with $\phi_s^L\sim-86^{\circ}$ are crucial to reconcile the puzzle. However, the $Z^{\prime}$ contributions involving $\zeta^{LR}_s$, which has a large uncertainty, are found to be almost irrelevant. It is interesting to investigate if further information on the model parameters could be obtained from the current experimental data on tree-dominated $B\to\pi\pi$ decays, which is the motivation of Case~III.

\begin{figure}[t]
\begin{center}
\epsfxsize=15cm \centerline{\epsffile{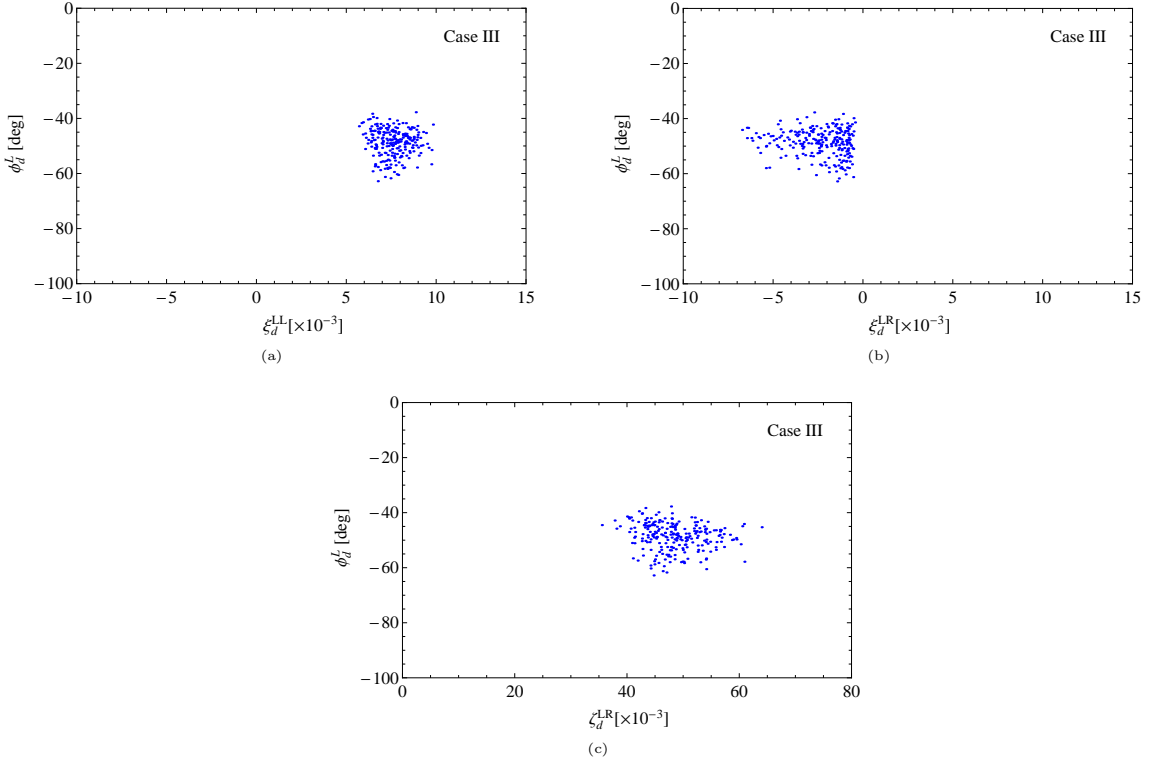}}
\centerline{\parbox{16cm}{\caption{\label{CaseIII}\small (Color online) The final allowed regions for the $Z^{\prime}$ coupling parameters $\xi^{LL}_d$, $\xi^{LR}_d$, $\zeta^{LR}_d$ and $\phi^L_d$ in Case~III. }}}
\end{center}
\end{figure}

As discussed in Ref.~\cite{Chang}, the ratios $R_{1,3}$ defined by Eq.~(\ref{Ratio}) have already been severely constrained, with
%%%%%%%%%%%%%%%%%%%%%%%%%%%%%%%%%%%%%%%%%%%
\begin{equation}\label{R13previous}
-0.77\leqslant R_{1}\leqslant-0.07\,, \qquad -3.7\leqslant R_{3}\leqslant6.5\,.
\end{equation}
%%%%%%%%%%%%%%%%%%%%%%%%%%%%%%%%%%%%%%%%%%%
Taking these ranges as input, we are then left with only two free parameters $\phi_d^L$ and $\xi^{LL}_d$~\footnote{Of course one could choose any one of $\xi^{LL}_d$, $\xi^{LR}_d$ and $\zeta^{LR}_d$ as the free parameter, and the other two can then be reduced from Eq.~(\ref{Ratio}). It is found that these different choices are irrelevant to our final fitting results.}. With the current experimental data on ${\cal B}(B\to\pi\pi)$ and $A_{CP}(B\to\pi\pi)$, as well as the ranges for $R_{1,3}$ given by Eq.~(\ref{R13previous}) as constraints, the final allowed regions for the parameters $\xi^{LL}_d$, $\xi^{LR}_d$, $\zeta^{LR}_d$ and $\phi^L_d$ are shown in Fig.~\ref{CaseIII}. The corresponding numerical results for these parameters are listed in the last column of Table~\ref{NPPareValue}. As the constraints Eq.~(\ref{R13previous}) are obtained with the regulation scheme~II~\cite{Chang}, for consistency, in this case we just use regulation scheme~II for the end-point divergency in our numerical evaluations.

Comparing Fig.~\ref{CaseIII} with Fig.~\ref{CaseII}, one can see that a large part of the parameter spaces of $\xi^{LL,LR}_d$ allowed in Case~II is excluded by the ratios $R_{1,3}$. However, the effects of $R_{1,3}$ on $\zeta^{LL}_d$ are found to be very tiny. On the other hand, the data on $B\to\pi\pi$ decays could put a   further constraint on the ratio $R_{3}$, $4.5\leqslant R_{3}\leqslant6.5 $, which implies that the part of parameter spaces with $\zeta^{LR}_s<4.5\times\xi^{LL}_s$ obtained in Ref.~\cite{Chang} will be excluded. The range of the ratio $R_{1}$ remains almost unchanged. As a consequence, the NP parameter spaces are severely reduced but still not excluded entirely, which means that both the ``$\pi K$'' and ``$\pi\pi$'' puzzles could be accommodated simultaneously within such a specific model.

%%%%%%%%%%%%%%%%%%%%%%%%%%%%%%%%%%%%%%%%
%%%%%%%%%%%%%%%%%%%%%%%%%%%%%%%%%%%%%%%%%%%

In order to see the relative strength of the flavour-changing couplings $B_{db}^L$ and $B_{sb}^L$, we define  the ratio
%%%%%%%%%%%%%%%%%%%%%%%%%%%%%%%%%%%%%%%%%%%%%%%%%
\begin{equation}\label{Rds}
 R_{ds}\,\equiv\,\frac{\xi^{LL,LR}_d}{\xi^{LL,LR}_s}\,=\,
 \big|\frac{V_{ts}^{\ast}}{V_{td}^{\ast}}\big|\,\big|\frac{B_{db}^L}{B_{sb}^L}\big|\,.
\end{equation}
%%%%%%%%%%%%%%%%%%%%%%%%%%%%%%%%%%%%%%%%%%%%%%%%%
With the central values of $\xi^{LL,LR}_d$ listed in Table~\ref{NPPareValue}, $\xi^{LL,LR}_s$ from
Ref.~\cite{Chang} and the CKM parameters listed in Appendix~C2, we get numerically
%%%%%%%%%%%%%%%%%%%%%%%%%%%%%%%%%%%%%%%%%%%%%%%%%
\begin{equation}
\frac{\xi^{LL}_d}{\xi^{LL}_s}=4.6,\qquad\frac{\xi^{LR}_d}{\xi^{LR}_s}=4.5,\qquad
|\frac{V_{ts}^{\ast}}{V_{td}^{\ast}}|=4.8\,,
\end{equation}
%%%%%%%%%%%%%%%%%%%%%%%%%%%%%%%%%%%%%%%%%%%%%%%%%
which implies an interesting relation $|B_{db}^L|\simeq |B_{sb}^L|$. Thus, in such a family non-universal $Z^{\prime}$ model, the difference of flavour-changing $Z^{\prime}$ couplings between the quark-level transitions $b\to d\bar{q}q$ and $b\to s\bar{q}q$ arises only from the new weak phase $\phi_{p}^L$, with $\phi_{d}^L\sim-50^{\circ}$ and $\phi_{s}^L\sim-85^{\circ}$~\cite{Chang}.

It should be noted that the $B_{d}^0-\bar{B}_{d}^0$ mixing puts a very strong constraint on the $b-d-Z^{\prime}$ coupling $B_{db}^{L}$~\cite{Liu1,Liu2,Chang:2010zy1,Chang:2010zy2,Chiang:2006we,Chang:2010Bs}. As the $Z^{\prime}$ contributions to $B\to\pi\pi$ decays involve not only the $b-d-Z^{\prime}$ coupling but also the flavor-conserving $Z^{\prime}$ couplings $d-d-Z^{\prime}$~($B_{dd}^{L,R}$) and $u-u-Z^{\prime}$~($B_{uu}^{L,R}$), one could not extract the constraint on $B_{db}^{L}$ solely from these decays. However, we find that our result $\phi_d^{L}\sim-50^{\circ}$ agrees well with the recent combined constraint $\phi_{d}^L=-33^{\circ}\pm45^{\circ}$~\cite{Chang:2010Bs} from $B_{d}^0-\bar{B}_{d}^0$ mixing and $\bar{B}_s\to\pi^- K^+$ decay.

Moreover, with the constraint from $B_{s}^0-\bar{B}_{s}^0$ mixing on the $b-s-Z^{\prime}$ coupling $B_{sb}^{L}$ included, it is found that $|B_{db}^{L}/B_{sb}^{L}|\sim\mathcal{O}(10^{-1})$~\cite{Chang:2010zy1,Chang:2010zy2}, which is inconsistent with the relation $|B_{db}^L|\simeq |B_{sb}^L|$ found in this paper. It is noted that the result $|B_{db}^{L}/B_{sb}^{L}|\sim\mathcal{O}(10^{-1})$ is based on the former CDF and D0 combined result~\cite{CDFD0old} for the $B_{s}^0-\bar{B}_{s}^0$ mixing, however, both the CDF and D0 collaborations have updated their measurements of the weak phase $\beta_s$ very recently,
\begin{equation}
\beta_s=\left\{\begin{array}{l}
[0.02,0.52]\cup[1.08,1.55]\,\quad {\rm CDF}~\cite{CDFnew}\,, \\
0.38^{+0.18}_{-0.19}\pm0.01\,\quad {\rm D0}~\cite{D0new},
\end{array} \right.
\end{equation}
which are different from the former CDF and D0 combined result $[0.27, 0.59]\cup[0.97, 1.30]$~\cite{CDFD0old}. Notably the CDF updated measurement~\cite{CDFnew} agrees with the SM expectation $\beta_s\sim0.018$ at $\sim1\sigma$ level, while the D0 updated measurement~\cite{D0new} agrees with their former combined result. As discussed in Ref.~\cite{Chang:2010Bs}, if the lower bound of the CDF updated measurement $\beta_s\sim0.02$ is allowed, one may easily find that the relation $|B_{db}^L|\simeq |B_{sb}^L|$ still survives the constraints from $B_{d,s}-\bar{B}_{d,s}$ mixings. In such a situation, refined experimental measurements and theoretical predictions are therefore expected urgently.

\section{Conclusions}

Motivated by the large discrepancy of the ratio $R_{00}^{\pi\pi}$ between theoretical predictions and experimental measurements, we have investigated the effect of a family non-universal $Z^{\prime}$ model on the tree-dominated $B\to \pi\pi$ decays, pursuing possible resolutions to the observed ``$\pi \pi$  puzzle''. Moreover, we have also taken into account the fitting results from the penguin-dominated $B\to \pi K$, $\pi K^{\ast}$ and $\rho K$ decays in Case~III, which gives a much stronger constraint on the flavour-changing $Z^{\prime}$ couplings. Our main conclusions are summarized as:

\begin{itemize}
\item The $Z^{\prime}$ contributions with $\zeta^{LR}_d\sim0.05$ and $\phi^{L}_d\sim-50^{\circ}$, being mainly relevant to the coefficient of QCD penguin operator $O_5$, are crucial to bridge the large discrepancy of ${\cal B}(\bar{B}^0\to\pi^0 \pi^0)$ between the theoretical prediction and the experimental measurement. The contributions involving the other parameters $\xi^{LL,LR}_d$ and $\zeta^{LL}_d$ are almost irrelevant to the observed ``$\pi\pi$ puzzle''.

\item Combining with the fitting results of $R_{1,3}$ from the penguin-dominated $B\to \pi K$, $\pi K^{\ast}$ and $\rho K$ decays, the NP parameter spaces are severely reduced but still not excluded entirely. This means that both the ``$\pi K$'' and ``$\pi\pi$'' puzzles could be accommodated simultaneously within such a family non-universal $Z^{\prime}$ model.

\item For all of the three different cases, a new weak phase $\phi_{d}^L$ associated with the flavour-changing $Z^\prime$ coupling $B_{db}^{L}$, with a value around $-50^{\circ}$, is always required for resolving the observed discrepancies.

\item The flavour-changing $Z^{\prime}$ couplings $|B_{db}^L|$ and $|B_{sb}^L|$, corresponding to the quark-level transitions $b\to d\bar{q}q$ and $b\to s\bar{q}q$ transitions, respectively, are found to be almost equal to each other.
\end{itemize}

With the upcoming LHC-b and proposed super-B experiments, the data on B-meson decays is expected to be more precise~\cite{newdata1,newdata2,newdata3}, which will then severely shrink or totally excluded the model parameter spaces. It is also reminded that more refined measurements of mix-induced CP asymmetries in these decays are urgently needed to further constrain the $Z^\prime$ coupling parameters.

\section*{Acknowledgments}
The work was supported in part by the National Natural Science Foundation under contract Nos.~11075059, 10735080 and 11005032.  X.~Q. Li was also supported in part by MEC (Spain) under Grant FPA2007-60323 and by the Spanish Consolider Ingenio 2010 Programme CPAN (CSD2007-00042).

%%%%%%%%%%%%%%%%%%%%%%%%%%%%%%%%%%%%%%%%%%%%
\begin{appendix}

\section*{Appendix~A: Decay amplitudes in the SM with QCDF}
The decay amplitudes for $B\to\pi\pi$ decays are recapitulated from Ref.~\cite{Beneke3}:
\begin{eqnarray}
\sqrt{2}{\cal A}_{B^-\to\pi^-\pi^0}^{\rm SM}
   &=&\sum_{p=u,c}V_{pb}V_{pd}^{\ast}A_{\pi\pi} \Big[
    \delta_{pu}\,(\alpha_1+\alpha_2)+\frac{3}{2}\,(\alpha_{3,{\rm
    EW}}^p+\alpha_{4,{\rm EW}}^p)\,\Big]\,,
\label{amp1_SM}\\[0.15cm]
{\cal A}_{\bar{B}^0\to\pi^+ \pi^-}^{\rm SM}
   &=& \sum_{p=u,c}V_{pb}V_{pd}^{\ast}A_{\pi \pi} \Big[
    \delta_{pu}\,(\alpha_1+\beta_1)+ \alpha_{4}^p+\alpha_{4,{\rm
    EW}}^p+\beta_3^p+2\,\beta_4^p\nonumber\\
    &&-\half\,(\beta_{3,{\rm EW}}^p-\beta_{4,{\rm EW}}^p)\,\Big]\,,
\label{amp2_SM}\\[0.15cm]
 -{\cal A}_{\bar{B}^0\to\pi^0 \pi^0}^{\rm SM}
   &=& \sum_{p=u,c}V_{pb}V_{pd}^{\ast}A_{\pi \pi} \Big[
    \delta_{pu}\,(\alpha_2-\beta_1)- \alpha_{4}^p+\frac{3}{2}\alpha_{3,{\rm
    EW}}^p+\half\alpha_{4,{\rm
    EW}}^p-\beta_3^p-2\,\beta_4^p\nonumber\\
    &&+\half\,(\beta_{3,{\rm EW}}^p-\beta_{4,{\rm EW}}^p)\,\Big]\,,
\label{amp3_SM}
\end{eqnarray}
where explicit expressions for the effective coefficients $\alpha_i^p\equiv\alpha_i^p(M_1M_2)$ and $\beta_i^p\equiv\beta_i^p(M_1M_2)$ could also be found in Ref.~\cite{Beneke3}. It should be noted that the hard-spectator terms $H_i$ appearing in $\alpha_i^p$ and the weak annihilation terms $A_{j}^{i,f}$ appearing in $\beta_j^p$ should be replaced by our recalculated ones listed in Appendix B.

\section*{Appendix~B: The hard-spectator and annihilation kernels with the infrared finite gluon propagator}
With the infrared finite gluon propagator to cure the end-point divergence, the hard-spectator kernels
in a general $B\to PP$ decay can be expressed as~\cite{YDYang}
%%%%%%%%%%%%%%%%%%%%%%%%%%%%%%%%%%%%%%%%%%%%%%%%%
\begin{equation}
H_i(M_1M_2)= \frac{B_{M_1 M_2}}{A_{M_1 M_2}} \int_0^1dxdyd\xi
\frac{\alpha_s(q^2)}{\xi}\Phi_{B1}(\xi)\Phi_{M_2}(x)\Big[\frac{\Phi_{M_1}(y)}
{\bar{x}(\bar{y}+\omega^2(q^2)/\xi)}+r_\chi^{M_1}\frac{\phi_{m_1}(y)}
{x(\bar{y}+\omega^2(q^2)/\xi)}\Big]\,,
\label{hard1}
\end{equation}
%%%%%%%%%%%%%%%%%%%%%%%%%%%%%%%%%%%%%%%%%%%%%%%%%
for the insertion of operators $Q_{i=1-4,9,10}$,
%%%%%%%%%%%%%%%%%%%%%%%%%%%%%%%%%%%%%%%%%%%%%%%%%
\begin{equation}
H_i(M_1M_2)= -\frac{B_{M_1 M_2}}{A_{M_1 M_2}} \int_0^1dxdyd\xi
\frac{\alpha_s(q^2)}{\xi}\Phi_{B1}(\xi)\Phi_{M_2}(x)\Big[\frac{\Phi_{M_1}(y)}
{x(\bar{y}+\omega^2(q^2)/\xi)}+r_\chi^{M_1}\frac{\phi_{m_1}(y)}{\bar{x}
(\bar{y}+\omega^2(q^2)/\xi)}\Big], \label{hard2}
\end{equation}
%%%%%%%%%%%%%%%%%%%%%%%%%%%%%%%%%%%%%%%%%%%%%%%%%
for $Q_{i=5,7}$, and $H_i(M_1M_2)=0$ for $Q_{i=6,8}$. When both $M_1$ and $M_2$ are pseudo-scalars,
the final building blocks for annihilation contributions can be expressed as~\cite{YDYang}
%%%%%%%%%%%%%%%%%%%%%%%%%%%%%%%%%%%%%%%%%%%%%%%%%
\begin{eqnarray}
A_1^i&=&\pi\int_0^1dxdy\alpha_s(q^2)\biggl\{\Big[
\frac{\bar{x}}{(\bar{x}y-\omega^2(q^2)+i\epsilon)(1-x\bar{y})}+
\frac{1}{(\bar{x}y-\omega^2(q^2)+i\epsilon)\bar{x}}\Big]\Phi_{M_1}(y)\Phi_{M_2}(x)
\nonumber\\[0.2cm]
&&+\frac{2}{\bar{x}y-\omega^2(q^2)+i\epsilon}r_\chi^{M_1}r_\chi^{M_2}\phi_{m_1}
(y)\phi_{m_2}(x)\biggl\}~,\label{anni1}\\[0.2cm]
A_2^i&=&\pi\int_0^1dxdy\alpha_s(q^2)\biggl\{\Big[
\frac{y}{(\bar{x}y-\omega^2(q^2)+i\epsilon)(1-x\bar{y})}+
\frac{1}{(\bar{x}y-\omega^2(q^2)+i\epsilon)y}\Big]\Phi_{M_1}(y)\Phi_{M_2}(x)
\nonumber\\[0.2cm]
&&+\frac{2}{\bar{x}y-\omega^2(q^2)+i\epsilon}r_\chi^{M_1}r_\chi^{M_2}\phi_{m_1}(y)
\phi_{m_2}(x)\biggl\},~\label{anni2}\\[0.2cm]
A_3^i&=&\pi\int_0^1dxdy\alpha_s(q^2)\biggl\{\frac{2\bar{y}}{(\bar{x}y-\omega^2(q^2)
+i\epsilon)(1-x\bar{y})}
r_\chi^{M_1}\phi_{m_1}(y)\Phi_{M_2}(x)\nonumber\\[0.2cm]
&&-\frac{2x}{(\bar{x}y-\omega^2(q^2)+i\epsilon)(1-x\bar{y})}r_\chi^{M_2}(x)
\phi_{m_2}(x)\Phi_{M_1}(y)\biggl\}~,\label{anni3}
\end{eqnarray}
\begin{eqnarray}
A_1^f&=&A_2^f=0,~\label{anni4}\\[0.2cm]
A_3^f&=&\pi\int_0^1dxdy\alpha_s(q^2)\biggl\{\frac{2(1+\bar{x})}{(\bar{x}y-
\omega^2(q^2)+i\epsilon)\bar{x}}
r_\chi^{M_1}\phi_{m_1}(y)\Phi_{M_2}(x)\nonumber\\[0.2cm]
&&+\frac{2(1+y)}{(\bar{x}y-\omega^2(q^2)+i\epsilon)y}r_\chi^{M_2}(x)\phi_{m_2}(x)
\Phi_{M_1}(y)\biggl\}~.\label{anni5}
\end{eqnarray}
%%%%%%%%%%%%%%%%%%%%%%%%%%%%%%%%%%%%%%%%%%%%%%%%%

\section*{Appendix~C: Theoretical input parameters}

\subsection*{C1. The numerical results of Wilson coefficients}
The numerical results of Wilson coefficients in the naive dimensional regularization~(NDR) scheme at
the scale $\mu=m_{b}$~($\mu_h=\sqrt{\Lambda_h m_b}$) are listed in Table~\ref{Wilson}. For simplicity,
we have defined
%%%%%%%%%%%%%%%%%%%%%%%%%%%%%%%%%%%%%%%%%%%%%%%%%
\begin{eqnarray}
 X=-\frac{|V_{tb}V_{td}^*|}{V_{tb}V_{td}^*}\xi^{LL}e^{i\phi_L}\,,\qquad
 Y=-\frac{|V_{tb}V_{td}^*|}{V_{tb}V_{td}^*}\xi^{LR}e^{i\phi_L}\,,\nonumber\\
 X^{\prime}=-\frac{|V_{tb}V_{td}^*|}{V_{tb}V_{td}^*}\zeta^{LL}e^{i\phi_L}\,,\qquad
 Y^{\prime}=-\frac{|V_{tb}V_{td}^*|}{V_{tb}V_{td}^*}\zeta^{LR}e^{i\phi_L}\,.
\end{eqnarray}
The values at the scale $\mu_{h}$, with $m_{b}=4.79~{\rm GeV}$ and $\Lambda_{h}=500~{\rm MeV}$, should
be used in the calculation of hard-spectator and annihilation contributions.

%%%%%%%%%%%%%Table wilson coefficients%%%%%%%%%%%%%%%%%%%%%%%%%%%
\begin{table}[t]
 \begin{center}
 \caption{\small The Wilson coefficients $C_i$ within the SM and the $Z^{\prime}$ model in NDR scheme at the scale $\mu=m_{b}$ and $\mu_h=\sqrt{\Lambda_h m_b}$, respectively.}
 \label{Wilson}
 \vspace{0.5cm}
 \small
 \doublerulesep 0.7pt \tabcolsep 0.03in
 \begin{tabular}{c|rr|rr}\hline\hline
 Wilson                   &\multicolumn{2}{c|}{$\mu=m_{b}$}                        &\multicolumn{2}{c}{$\mu_h=\sqrt{\Lambda_h m_b}$} \\\cline{2-3}\cline{4-5}
 coefficients             &$C_i^{SM}$ &$\Delta C_i^{Z^{\prime}}$                   &$C_i^{SM}$ &$\Delta C_i^{Z^{\prime}}$\\ \hline\hline
 $C_1$                    &$1.075$    &$-0.006X$                                   &$1.166$    &$-0.008X$\\
 $C_2$                    &$-0.170$   &$-0.009X$                                   &$-0.336$   &$-0.014X$ \\
 $C_3$                    &$0.013$    &$0.05X-0.01Y-2.20X^{\prime}-0.05Y^{\prime}$ &$0.025$    &$0.11X-0.02Y-2.37X^{\prime}-0.12Y^{\prime}$\\
 $C_4$                    &$-0.033$   &$-0.13X+0.01Y+0.55X^{\prime}+0.02Y^{\prime}$&$-0.057$   &$-0.24X+0.02Y+0.92X^{\prime}+0.09Y^{\prime}$\\
 $C_5$                    &$0.008$    &$0.03X+0.01Y-0.06X^{\prime}-1.83Y^{\prime}$ &$0.011$    &$0.03X+0.02Y-0.10X^{\prime}+0.09Y^{\prime}$\\
 $C_6$                    &$-0.038$   &$-0.15X+0.01Y+0.1X^{\prime}-0.6Y^{\prime}$  &$-0.076$   &$-0.32X+0.04Y+0.16X^{\prime}-1.26Y^{\prime}$\\
 $C_7/{\alpha}_{em}$      &$-0.015$   &$4.18X-473Y+0.25X^{\prime}+1.27Y^{\prime}$  &$-0.034$   &$5.7X-459Y+0.4X^{\prime}+1.7Y^{\prime}$\\
 $C_8/{\alpha}_{em}$      &$0.045$    &$1.18X-166Y+0.01X^{\prime}+0.56Y^{\prime}$  &$0.089$    &$3.2X-355Y+0.2X^{\prime}+1.5Y^{\prime}$\\
 $C_9/{\alpha}_{em}$      &$-1.119$   &$-561X+4.52Y-0.8X^{\prime}+0.4Y^{\prime}$   &$-1.228$   &$-611X+6.7Y-1.2X^{\prime}+0.6Y^{\prime}$\\
 $C_{10}/{\alpha}_{em}$   &$0.190$    &$118X-0.5Y+0.2X^{\prime}-0.05Y^{\prime}$    &$0.356$    &$207X-1.4Y+0.5X^{\prime}-0.1Y^{\prime}$\\
 $C_{7{\gamma}}$          &$-0.297$   &---                                         &$0.360$    &---\\
 $C_{8g}$                 &$-0.143$   &---                                         &$-0.168$   &---\\
 \hline \hline
 \end{tabular}
 \end{center}
 \end{table}
%%%%%%%%%%%%%Table wilson coefficients%%%%%%%%%%%%%%%%%%%%%%%%%%%

\subsection*{C2. CKM matrix elements}
For the CKM matrix elements, we adopt the Wolfenstein parameterization~\cite{Wolfenstein:1983yz} and choose the four parameters $A$, $\lambda$, $\rho$ and $\eta$ as~\cite{Charles:2004jd}
\begin{equation}
A=0.798^{+0.023}_{-0.017}, \qquad
\lambda=0.22521^{+0.00083}_{-0.00082}, \qquad
\overline{\rho}=0.141^{+0.035}_{-0.021}, \qquad
\overline{\eta}=0.340\pm0.016,
\end{equation}
with $\overline{\rho}=\rho\,(1-\frac{\lambda^2}{2})$ and
$\bar{\eta}=\eta\,(1-\frac{\lambda^2}{2})$.

\subsection*{C3. Quark masses and lifetimes}
As for the quark masses, there are two different classes appearing in our calculation. One type is the current quark mass which appears in the factor $r_\chi^M$ through the equation of motion for quarks.
This type of quark masses is scale dependent and denoted by $\overline{m}_q$. Here we take
\begin{equation}
\overline{m}_s(\mu)/\overline{m}_q(\mu)=27.4\pm0.4\,~\cite{HPQCD:2006},\quad
\overline{m}_{s}(2\,{\rm GeV}) =87\pm6\,{\rm
MeV}\,~\cite{HPQCD:2006}, \quad
\overline{m}_{b}(\overline{m}_{b})=4.20^{+0.17}_{-0.07}\,{\rm
GeV}~\cite{PDG08}\,,
\end{equation}
where $\overline{m}_q(\mu)=(\overline{m}_u+\overline{m}_d)(\mu)/2$, and the difference between $u$ and $d$ quark is not distinguished.

The other one is the pole quark mass appearing in the evaluation of penguin loop diagrams, and denoted by $m_q$. In this paper, we take
\begin{equation}
 m_u=m_d=m_s=0, \qquad m_c=1.61^{+0.08}_{-0.12}\,{\rm GeV}, \qquad
 m_b=4.79^{+0.19}_{-0.08}\,{\rm GeV}.
\end{equation}

As for the B-meson lifetimes, we take~\cite{PDG08} $\tau_{B_{u}} = 1.638\,{\rm ps}$ and $\tau_{B_{d}}=1.530\,{\rm ps}$, respectively.

\subsection*{C4. Input parameters related to mesons.}
In this paper, we take the heavy-to-light transition form factors
and the decay constants as
%%%%%%%%%%%%%%%%%%%%%%%%%%%%%%%%%
\begin{equation}
F^{B\to \pi}_{0}(0)=0.258\pm0.031\,~\cite{BallZwicky1,BallZwicky2,BallZwicky3}, \quad
f_{B}=(216\pm22)~{\rm MeV}~\cite{Gray:2005ad}\,, \quad
f_\pi=(130.4\pm0.2)~{\rm MeV}~\cite{PDG08}\,.
\end{equation}
%%%%%%%%%%%%%%%%%%%%%%%%%%%%%%%%

The light-cone projector operator of light pseudo-scalars in momentum space reads~\cite{Beneke3,Terentev}
%%%%%%%%%%%%%%%%%%%%%%%%%%%%%%%%%%%%%%%%%%%%
\begin{equation}\label{pimeson3}
   M_{\alpha\beta}^P = \frac{i f_P}{4} \left[
   \pslash\,\gamma_5\,\Phi_P(x) - \mu_P\gamma_5\,
   \frac{\kslash_2\,\kslash_1}{k_2\cdot k_1}\,\phi_p(x)
   \right]_{\alpha\beta}\,,
\end{equation}
%%%%%%%%%%%%%%%%%%%%%%%%%%%%%%%%%%%%%%%%%%%%
where $f_{P}$ is the decay constants, and $\mu_P=m_b r_\chi^P/2$ with the chirally-enhanced factor
$r_\chi^P$ defined as
%%%%%%%%%%%%%%%%%%%%%%%%%%%%%%%%%%%%%%%%%%%%
\begin{equation}
r_{\chi}^{\pi}(\mu)=\frac{2m_{\pi}^2}{\overline{m}_b(\mu)\,2\,\overline{m}_{q}(\mu)}\,,
\end{equation}
%%%%%%%%%%%%%%%%%%%%%%%%%%%%%%%%%%%%%%%%%%%%
where the quark masses are all running masses defined in the $\overline{\rm MS}$ scheme. For the light-cone distribution amplitude of light pseudo-scalars, we use their asymptotic forms~\cite{projector,formfactor1,formfactor2,formfactor3}
\begin{equation}
\Phi_{P}(x)=6\,x(1-x)\,, \quad \phi_p (x)=1\,.
\end{equation}

As for the B-meson wave function, we take the form~\cite{YYPhiB}
\begin{equation}
\Phi_B(\xi)=N_B\xi(1-\xi)\textmd{exp}\Big[-\Big(\frac{M_B}{M_B-m_b}\Big)^2(\xi-\xi_B)^2\Big],
\end{equation}
where $\xi_B\equiv1-m_b/M_B$, and $N_B$ is the normalization constant to insure that $\int_0^1 d\xi\Phi_B(\xi)=1$.

\end{appendix}

%%%%%%%%%%%%%%%%%%%%%%YD,PikPQCD,PikSCET,Pik%%%%%%%%%%%%%%%%%%%%%%

\end{document}